\documentclass[conference]{IEEEtran}
\IEEEoverridecommandlockouts
\usepackage{cite}
\usepackage{adjustbox}
\usepackage[ruled,vlined]{algorithm2e}
\usepackage{amsmath,amssymb,amsfonts}
\usepackage{algorithmic}
\usepackage{graphicx}
\usepackage{textcomp}
\usepackage[colorlinks=false, pdfborder={0 0 0}, breaklinks]{hyperref}
\usepackage{xcolor}
\usepackage{balance}
\usepackage{enumitem}
\usepackage{setspace}
\usepackage{wrapfig}
\usepackage{listings}
\usepackage{color}
\usepackage{subcaption}
\usepackage{chemmacros}
\usepackage[longtable]{multirow}

\usepackage{pifont}
\usepackage{xcolor}
\usepackage{tikz}
\usepackage{verbatim} 
\usepackage[accsupp]{axessibility}

\newcommand*\circled[1]{\tikz[baseline=(char.base)]{%
            \node[shape=circle,fill=green!20,draw,inner sep=0.5pt] (char) {#1};}}
            
\def\BibTeX{{\rm B\kern-.05em{\sc i\kern-.025em b}\kern-.08em
    T\kern-.1667em\lower.7ex\hbox{E}\kern-.125emX}}

\usepackage{amsmath}

\usepackage{mathtools}

\usepackage{tikz}
\usetikzlibrary{shapes.geometric, arrows}
\usetikzlibrary{fit}
\usetikzlibrary{calc}
\usetikzlibrary{shapes.geometric, arrows}
\tikzstyle{arrow} = [thick,->,>=stealth]

\usepackage{ctable}

\newcommand{\eg}{\emph{e.g.}, }

\newcommand{\HEVC}{\emph{High Efficiency Video Coding}\xspace}
\newcommand{\AVC}{\emph{Advanced Video Coding}\xspace}
\newcommand{\VVC}{\emph{Versatile Video Coding}\xspace}

\newcommand{\EY}{$E_{\text{Y}}$}
\newcommand{\LY}{$L_{\text{Y}}$}
\newcommand{\EU}{$E_{\text{U}}$}
\newcommand{\LU}{$L_{\text{U}}$}
\newcommand{\EV}{$E_{\text{V}}$}
\newcommand{\LV}{$L_{\text{V}}$}
\newcommand{\h}{$h$}

\newcommand{\vig}[1]{\textcolor{black}{#1}}

\title{Content-Driven Frame-Level Bit Prediction for Rate Control in Versatile Video Coding}

\author{\IEEEauthorblockN{Amritha Premkumar\IEEEauthorrefmark{1}, Prajit T Rajendran\IEEEauthorrefmark{2}, Vignesh V Menon\IEEEauthorrefmark{1}\IEEEauthorrefmark{3}, Christian Herglotz\IEEEauthorrefmark{1}
}
\IEEEauthorblockA{\IEEEauthorrefmark{1}Chair of Computer Engineering, Brandenburg University of Technology Cottbus-Senftenberg, Germany\\
\IEEEauthorrefmark{2}Universite Paris-Saclay, CEA, List, F-91120, Palaiseau, France\\
\IEEEauthorrefmark{3}Video Communication and Applications Dept, Fraunhofer HHI, Berlin, Germany
}
}

\begin{document}
\maketitle

\begin{abstract}
Rate control allocates bits efficiently across frames to meet a target bitrate while maintaining quality. Conventional two-pass rate control (2pRC) in Versatile Video Coding (VVC) relies on analytical rate–QP models, which often fail to capture nonlinear spatial–temporal variations, causing quality instability and high complexity due to multiple trial encodes. This paper proposes a content-adaptive framework that predicts frame-level bit consumption using lightweight features from the Video Complexity Analyzer (VCA) and quantization parameters within a Random Forest regression. On ultra-high-definition sequences encoded with VVenC, the model achieves strong correlation with ground truth, yielding $R^2$ values of 0.93, 0.88, and 0.77 for I-, P-, and B-frames, respectively. Integrated into a rate-control loop, it achieves comparable coding efficiency to 2pRC while reducing total encoding time by 33.3\%. The results show that VCA-driven bit prediction provides a computationally efficient and accurate alternative to conventional rate–QP models.
\end{abstract}


\section{Introduction}
\label{sec:intro}
The demand for high-quality video content continues to rise as streaming platforms, mobile devices, and over-the-top (OTT) services expand~\cite{CiscoForecast}. Ultra-high-definition (UHD) formats, high frame rates, and immersive applications such as virtual reality further increase the complexity of video delivery systems. In this context, video coding standards such as \AVC~(AVC)~\cite{AVC}, \HEVC~(HEVC)~\cite{HEVC}, and \VVC~(VVC)~\cite{bross_overview_2021} play a central role in enabling efficient compression and transmission. One of the key challenges in video coding is \emph{rate control}, which allocates bits across frames and coding units to meet target bitrate constraints while preserving perceptual quality~\cite{Helmrich_rc_icip24}. Accurate bit prediction underpins stable rate control and efficient storage and transmission.

Traditional rate control algorithms typically rely on simplified analytical models that relate the quantization parameter (QP) to output bitrate~\cite{Helmrich_rc_icip24}. These models often assume a linear or exponential relationship between QP and bits; however, actual bit consumption depends strongly on content characteristics. Classical rate–distortion (R–D) modeling for rate control has explored parametric distributions and frame-layer models~\cite{Fei_avc_rc,wang_rc,Rezaei_rc_fuzzy}, as well as schemes tailored for real-time and low-delay operation~\cite{Li_rc,Sanz_rc,wang_rc1,Naveen_rc}. Since bit consumption varies widely with spatial detail and motion, heuristic models often fail to generalize and require conservative buffer control or trial encodes to avoid overshoot, increasing computational cost and energy use. Recent advances in two-pass and fast-first-pass designs mitigate these issues via down-sampling and consistent-quality control~\cite{shen_rc,Henkel_rc_subsample,zhou_rc}.

To address this limitation, research has increasingly focused on content-adaptive rate control. Complexity measures such as Spatial Information (SI) and Temporal Information (TI), recommended by ITU, have been explored for predicting encoding behavior~\cite{siti_itu_ref}. However, their correlation with encoding outcomes such as bitrate or coding time is limited, especially for modern codecs~\cite{vca_ref}. More recently, machine learning methods have been applied to model the relationship between content features and encoding parameters, \vig{enabling content-aware two-pass optimization for bitrate ladder construction, per-title encoding, and rate control}~\cite{jtps_ref, azimi_decoding_2024}.

\begin{figure}[t]
\centering
  \includegraphics[width=0.4875\textwidth]{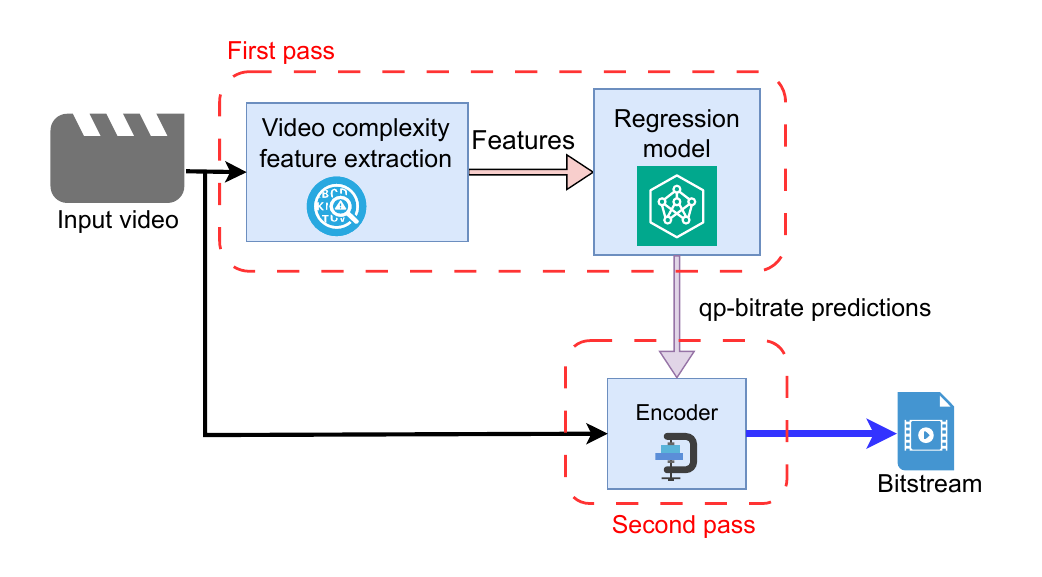}
\vspace{-2.22em}
\caption{Overview of the proposed two-pass rate control framework integrating VCA-driven bit prediction into VVenC. The first pass extracts content features, predicts frame-level bits, and refines QP assignment for the second-pass encoding.}
\vspace{-1.12em}
\label{fig:method}
\end{figure}

\vig{In contrast to recent learning-based rate control approaches that rely on encoder-internal features or deep networks, or approaches that focus on optimizing encoding resolution, bitrate ladders, or energy efficiency at the segment level~\cite{ladre_ref, katsenou2026multiobjectiveparetofrontoptimizationefficient},} this paper, as illustrated in Fig.~\ref{fig:method}, employs lightweight content features derived from the open-source Video Complexity Analyzer (VCA)~\cite{vca_ref} to predict frame-level bit consumption. VCA quantifies spatial texture energy, brightness, and inter-frame variations using frequency-domain descriptors derived from block-level DCT analysis. These features are computationally lightweight and correlate strongly with encoder decisions such as transform size and prediction mode selection, making them suitable for real-time bitrate estimation. Frame-type–specific models capture the distinct coding behavior of I-, P-, and B-frames, which differ in their dependence on spatial and temporal complexity. The proposed model directly estimates frame-level bit consumption from VCA features and QP during the first pass, enabling informed rate allocation and buffer control without an additional analysis pass. This narrows the gap between one-pass and two-pass schemes, approaching 2pRC accuracy while preserving single-pass latency, and provides a practical integration path into VVenC’s rate control loop.

Our evaluation on a large UHD dataset encoded with VVenC~\cite{vvenc_ref} shows that the proposed models achieve high accuracy, with the coefficient of determination ($R^2$) values of \SI{0.93}{}, \SI{0.88}{}, and \SI{0.77}{} for I, P, and B frames, respectively, while the mean absolute percentage errors (MAPE) remain low across frame types, confirming that VCA features effectively capture the video complexity that drives bitrate consumption.

The contributions of this paper are as follows:
\begin{enumerate}[topsep=0pt,leftmargin=*,label=\protect\circled{\arabic*}]
    \item A machine learning framework for predicting frame-level bit consumption in I, P, and B frames using VCA-derived features and QP.
    \item Validation that spatial and temporal complexity features guide accurate bit allocation, enabling energy-efficient and stable rate control in modern encoders.
\end{enumerate}

\section{Proposed Two-Pass Rate Control Framework}
\label{sec:firstpass}
\subsection{Video Complexity Feature Extraction}
\label{sec:feature_extraction}
The effectiveness of any machine learning-based bitrate prediction model depends strongly on the choice of input features. In this work, we employ seven features derived from the Video Complexity Analyzer (VCA)~\cite{vca_ref}, which extracts lightweight yet representative descriptors of spatial and temporal complexity, \{ \EY, \LY, \EU, \LU, \EV, \LV, \h \}, where $E$ quantifies spatial texture complexity, $L$ represents brightness, and $h$ measures inter-frame texture variation as a temporal complexity indicator~\cite{vca_ref}.

\begin{figure}[t]
\centering
  \includegraphics[width=0.4875\textwidth]{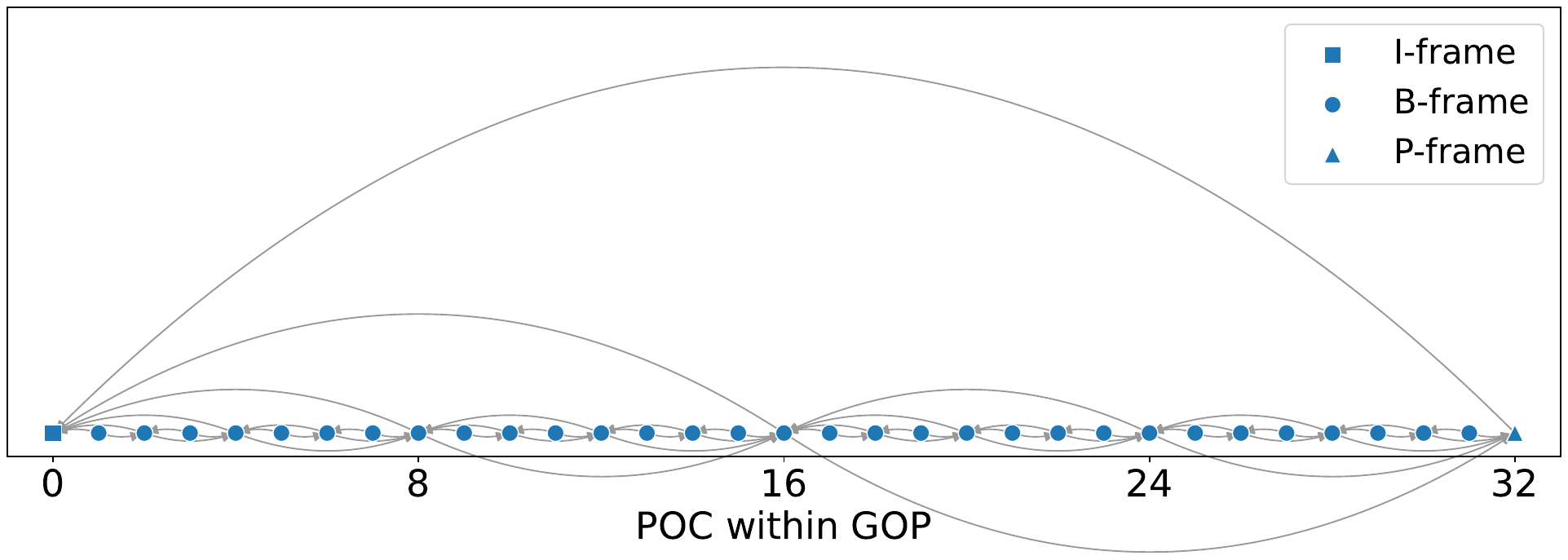}
\caption{Hierarchical GOP-32 structure showing inter-frame prediction dependencies (0-  I-frame, 32- P-frame, others- B-frames).}
\vspace{-1.2em}
\label{fig:gop}
\end{figure}

As illustrated in Fig.~\ref{fig:gop}, the Group of Pictures (GOP) structure follows a hierarchical bi-prediction pattern. To capture temporal dependencies consistent with this hierarchy, VCA is executed at multiple temporal sampling intervals. Feature extraction is performed for every frame and for subsampling configurations that analyze every 2, 4, 8, 16, or 32 frames, corresponding to temporal gaps of 1, 2, 4, 8, 16, and 32 frames. This multi-scale extraction captures both short- and long-term temporal dependencies across hierarchical levels. Features from smaller gaps emphasize rapid motion or scene changes, while larger gaps capture slow or global variations, improving generalization across diverse temporal dynamics.

\subsection{Regression Models}
\label{sec:prediction_models}

\subsubsection{I-frame}
I-frames are intra-coded without reference to other frames, and their bitrate depends primarily on spatial complexity~\cite{menon_AI_2pRC}. Hence, the number of bits to code an I-frame ($b_I$) is modeled as:
\begin{equation}
\label{eq:i_pred_model}
\hat{b}_{I} = f(E_{\text{Y}}, L_{\text{Y}}, E_{\text{U}}, L_{\text{U}}, E_{\text{V}}, L_{\text{V}}, q),
\end{equation}
where $b_I$ is represented as a function of its VCA-derived features and QP ($q$).

\subsubsection{P-frame}
P-frames exploit inter-frame prediction, referencing a past frame. Consequently, their bit consumption depends not only on spatial features but also on temporal features that capture motion and differences relative to the reference frame. Hence, the number of bits to code a P-frame ($b_P$) is modeled as:
\begin{equation}
\label{eq:p_pred_model}
\hat{b}_{P} = f(E_{\text{Y}}, h_{ref}, L_{\text{Y}}, E_{\text{U}}, L_{\text{U}}, E_{\text{V}}, L_{\text{V}}, q, q_{ref}),
\end{equation}
where, $h_{ref}$, which quantifies the difference in texture energy between the current and reference frames in the luma plane, and $q_{ref}$ denotes the QP of the reference frame.

\subsubsection{B-frame}
B-frames are bi-directionally predicted, referencing both past and future frames. Their bit consumption depends on a combination of spatial complexity, temporal gradients relative to two references ($h_{ref_{1}}$ and $h_{ref_{2}}$), and QPs of both the current and reference frames ($q_{ref_1}$ and $q_{ref_2}$), as illustrated in the equation below:
\begin{equation}
\label{eq:b_pred_model}
\begin{split}
\hat{b}_{B} = f(&E_{\text{Y}}, h_{ref_{1}}, h_{ref_{2}}, L_{\text{Y}}, E_{\text{U}}, L_{\text{U}}, \\
          &E_{\text{V}}, L_{\text{V}}, q, q_{ref_1}, q_{ref_2}).
\end{split}
\end{equation}
Although B-frames generally consume fewer bits than I- or P-frames, we expect their prediction to be more challenging due to multiple dependencies.

\subsection{Second Pass}
In the second encoding pass, the bit prediction from the first pass is utilized to refine the quantization parameter (QP) assignment for each frame. The first pass rate–QP estimator described in Section~\ref{sec:firstpass} provides an initial QP value $q$ and a corresponding bit prediction $\hat{b}$. Using this data pair and the target bit count $b'$ for the second pass, VVenC’s internal $R$–$QP$ model~\cite{rc_ref} determines the adjusted QP $q'$ that best meets the target rate.

As discussed in \cite{rc_ref}, the initial estimate $\bar{q}$ is computed as:
\begin{equation}
\bar{q} = q - c_\text{low} \cdot \sqrt{\max(1; q)} \cdot \log_2 \!\biggl(\frac{b'}{\hat{b}}\biggr),
\label{eq:eq1}
\end{equation}
where $c_\text{low}$ is a scaling constant associated with the low-rate region of the $R$–$QP$ curve. To improve robustness at higher bitrates, a corrective term is added:
\begin{equation}
q' = \text{round}\!\bigl(\bar{q} + c_\text{high} \cdot \max(0; q_\text{start} - \bar{q})\bigr),
\label{eq:eq12}
\end{equation}
where $c_\text{high}$ $(0 < c_\text{high} < 1)$ is a resolution-dependent constant (\eg 0.5 for 2160p and 0.25 for 480p content), and $q_\text{start}=24$ was empirically chosen as the reference point.

This procedure ensures that the predicted bit allocation is mapped to the most suitable QP value before the final rate–distortion optimized encoding. Any residual deviations between the achieved and target bit counts after this pass are compensated for adaptively in subsequent frames, as outlined in~\cite{rc_ref}. \vig{At the GOP level, the target bitrate is distributed proportionally across frames based on the predicted frame-level bit consumption $\hat{b}$ from the first pass, ensuring that relative complexity differences within the GOP are preserved while meeting the overall rate constraint.}, thereby providing a more reliable basis for downstream rate allocation decisions and helping maintain consistent coding behavior across challenging sequences exhibiting heterogeneous motion, texture variation, or abrupt scene transitions within the same GOP. 




\section{Evaluation Setup}
\label{sec:eval_setup}
\vspace{-0.3em}
\subsection{First Pass}
\subsubsection{Dataset}
For training the regression models in this study, we employ the Inter-4K dataset~\cite{stergiou_adapool_2023}, which contains 1000 diverse UHD video sequences and exposes the models to in-the-wild characteristics beyond the relatively stable scene structure of CTC test sequences. Videos were encoded using VVenC~\cite{vvenc_ref} at QPs $\in [20,50]$ using the RA coding structure~\cite{seq1_ref}, a GOP size of 32 frames, an intra period of 64 frames, and the \emph{faster} preset~\cite{preset_ref}. 
To verify model generalizability, 5-fold cross-validation is performed and results are averaged across all folds.  

\subsubsection{Feature Extraction}
Video complexity features are extracted using VCA version 1.5~\cite{vca_ref}. Block-level Discrete Cosine Transform (DCT) features are aggregated to frame-level descriptors. 
The temporal gradient feature $h$ is computed relative to one or two reference frames, depending on whether the current frame is a P- or B-frame. Extraction is performed with 8 CPU threads using x86 SIMD optimizations.

\subsubsection{Learning Framework}
We employ linear regression, random forest regression~\cite{breiman_random_2001}, and XGBoost~\cite{chen_xgboost_2016} for bit estimation. \vig{We choose these over recurrent or sequence-based models (\eg LSTM), as VCA features already encode temporal dependencies explicitly via multi-scale gradients. Moreover, tree-based models provide strong performance with significantly lower inference latency and memory footprint.} Training labels are obtained from the actual number of bits used by the encoder per frame. Features are normalized, and models are trained separately for I-, P-, and B-frames. The hyperparameters include the number of estimators (\texttt{n\_estimators}), maximum tree depth (\texttt{max\_depth}), and splitting and leaf node constraints. In our experiments, \texttt{n\_estimators} is fixed at 100, with \texttt{min\_samples\_split} and \texttt{min\_samples\_leaf} set to 2 and 1, respectively, and a maximum tree depth of 16. All models are implemented using \texttt{scikit-learn}~\cite{scikit_learn} with \texttt{random\_state=0}.

\subsubsection{Performance Metrics}
Prediction accuracy is assessed using:
\begin{itemize}[leftmargin=*,noitemsep,topsep=0pt]
    \item Mean Absolute Percentage Error (MAPE): Measures average prediction error magnitude in percent.
    \item $R^2$ score: Indicates the proportion of variance in the ground truth explained by the model.
\end{itemize}

\subsection{Second Pass}
RD performance is evaluated using the RA configuration at the \textit{faster} preset~\cite{preset_ref}, without temporal filtering. \vig{While absolute bit consumption characteristics change with slower presets, the proposed content–bit relationship is expected to remain valid due to its dependence on fundamental spatial–temporal features rather than preset-specific heuristics.} A predefined target rate is obtained from CTC-like coding with fixed QP. JVET SDR CTC sequences (classes A1 and A2)~\cite{seq1_ref} are used for evaluation and converted to 8-bit, 30 fps format. Experiments are conducted on an Intel Xeon E5-2697A v4 system running Linux and GCC 7.3.1 with eight CPU threads. The anchor is fixed-QP RA VVenC encoding using the \textit{faster} preset.

Results are reported using combined YUV Bjontegaard Delta-Rate ($BD_{YUV}$)~\cite{DCC_BJDelta}, average encoding time difference ($\Delta T$), and deviation from the target bitrate. Lower BD-rate indicates better coding efficiency at equivalent visual quality.

\section{Experimental Results}
\label{sec:evaluation}
\subsection{Prediction Accuracy}

\begin{figure}[t]
\centering
    \begin{subfigure}{0.235\textwidth}
    \centering
    \includegraphics[width=\textwidth]{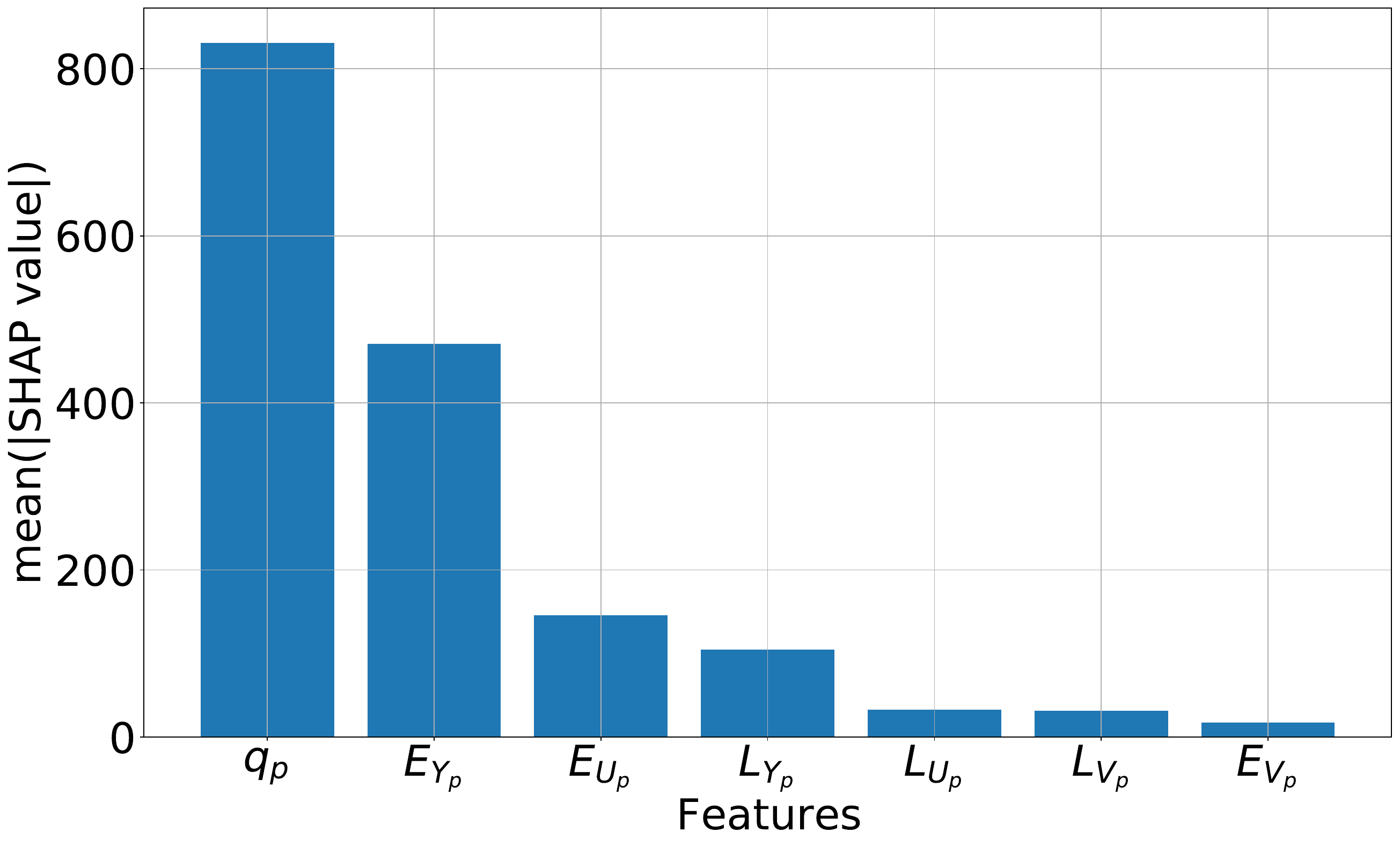}
    \caption{I-frame}
\label{fig:shap_i}
\end{subfigure}
\hfill
\begin{subfigure}{0.235\textwidth}
    \centering
    \includegraphics[width=\textwidth]{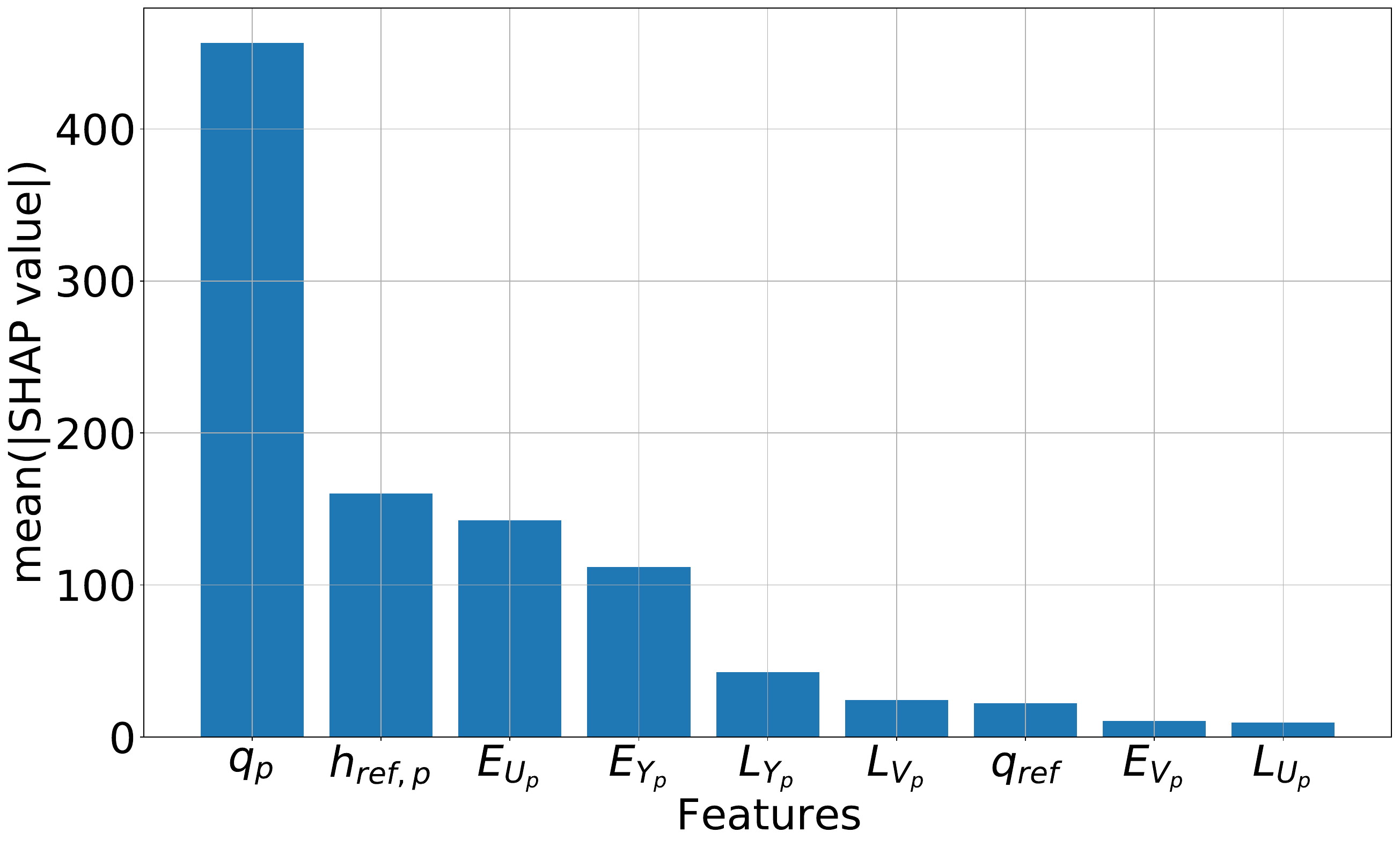}
    \caption{P-frame}
\label{fig:shap_p}
\end{subfigure}
\begin{subfigure}{0.40\textwidth}
    \centering
    \includegraphics[width=\textwidth]{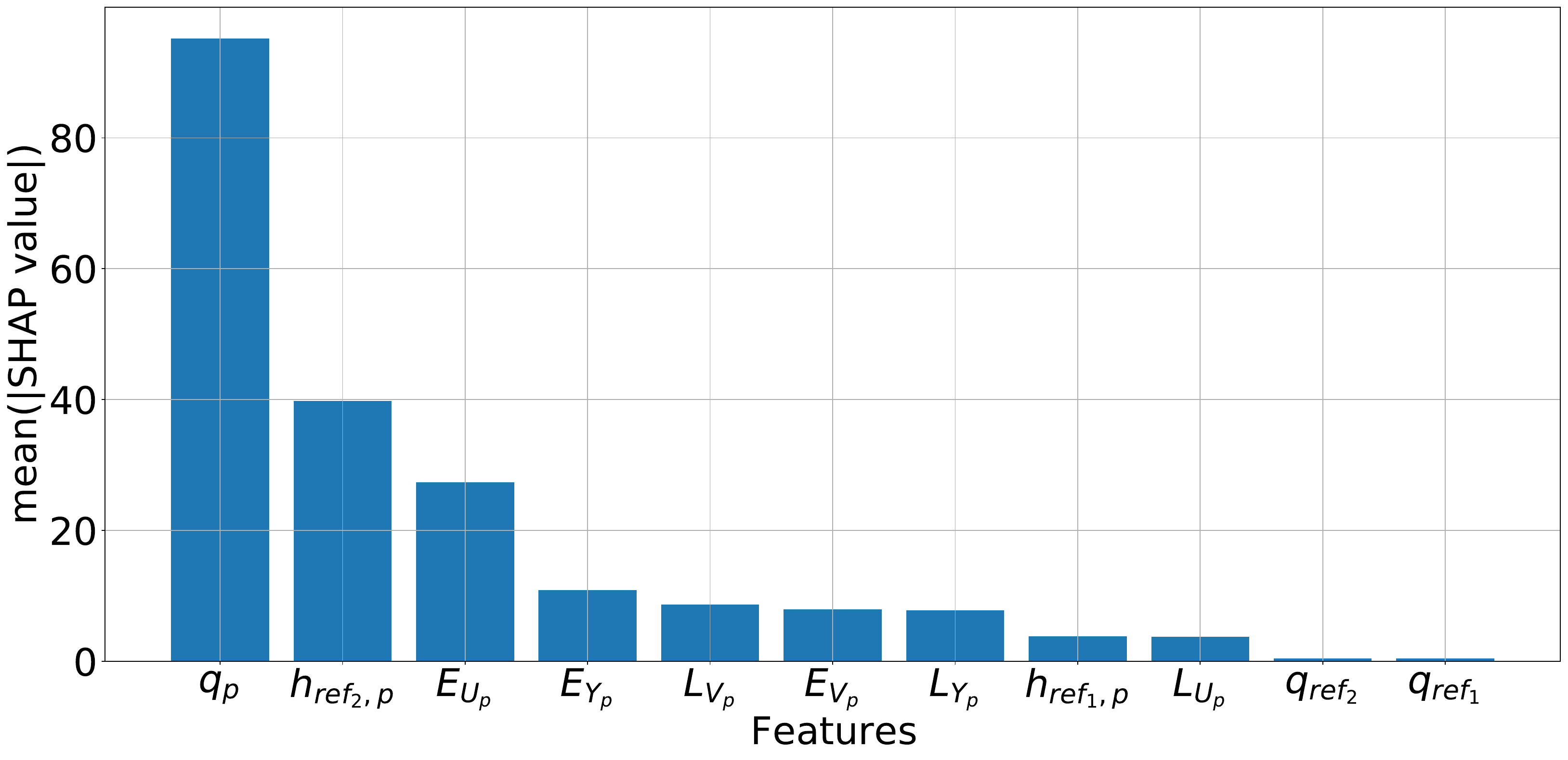}
    \caption{B-frame}
\label{fig:shap_b_double}
\end{subfigure}
\caption{Relative importance of features in bits prediction.}
\vspace{-1.8em}
\end{figure}

\subsubsection{Model Comparison and Ablation Study}
Table~\ref{tab:model_res_cons} summarizes the performance of different regression models for frame-level bit prediction across I-, P-, and B-frames. Among the evaluated models, Random Forest (RF) consistently achieves the best performance, yielding $R^2$ values of \SI{0.93}{}, \SI{0.88}{}, and \SI{0.77}{} for I, P, and B-frames, respectively; hence, RF is used for our rate control evaluation presented in Section~\ref{sec:ratecontrol_implications}. Its prediction errors remain below \SI{9}{\percent} MAPE across all frame types, indicating strong generalization and stable learning.

\subsubsection{I-frame}
Fig.~\ref{fig:shap_i} illustrates the relative importance of features in bit prediction, measured using SHAP values~\cite{shap_ref}. The quantization parameter $q$ dominates the ranking, reflecting its direct control over bit allocation in intra frames. The next most influential descriptor is the luma texture energy $E_{\text{Y}}$, which captures spatial frequency characteristics of the frame. Chroma texture energies ($E_{\text{U}}$, $E_{\text{V}}$) and luminance terms ($L_{\text{Y}}$, $L_{\text{U}}$, $L_{\text{V}}$) contribute less but remain useful for fine-grained prediction. For I-frames, the Random Forest model achieves a MAPE of \SI{6.84}{\percent} and $R^2=0.93$, indicating robustness.

\begin{table}[t]
\caption{Performance of frame-level bit prediction models.}
\centering
\resizebox{0.995\linewidth}{!}{
\begin{tabular}{l|c|c|c|c|c|c}
\specialrule{.12em}{.05em}{.05em}
\specialrule{.12em}{.05em}{.05em}
& \multicolumn{2}{c|}{I-frame} & \multicolumn{2}{c|}{P-frame} & \multicolumn{2}{c}{B-frame} \\
Model & MAPE & $R^2$ & MAPE & $R^2$ & MAPE & $R^2$ \\
& [\%] & & [\%] & & [\%] & \\
\specialrule{.12em}{.05em}{.05em}
\specialrule{.12em}{.05em}{.05em}
Linear regression [no chroma] & 8.92 & 0.86 & 11.15 & 0.79 & 17.48 & 0.65 \\
XGBoost [no chroma]            & 7.81 & 0.89 & 10.04 & 0.83 & 15.72 & 0.70 \\
Random forest [no chroma]      & 7.35 & 0.90 & 9.46  & 0.84 & 14.85 & 0.72 \\
\specialrule{.08em}{.05em}{.05em}
Linear regression              & 8.51 & 0.87 & 10.72 & 0.80 & 16.82 & 0.67 \\
XGBoost                        & 7.12 & 0.91 & 8.83  & 0.86 & 13.26 & 0.74 \\
Random forest                  & \textbf{6.84} & \textbf{0.93} & \textbf{8.21} & \textbf{0.88} & \textbf{12.94} & \textbf{0.77} \\
\specialrule{.12em}{.05em}{.05em}
\specialrule{.12em}{.05em}{.05em}
\end{tabular}}
\vspace{-0.7em}
\label{tab:model_res_cons}
\end{table}

\subsubsection{P-frame}
P-frame prediction integrates both spatial and temporal information. As shown in Fig.~\ref{fig:shap_p}, temporal gradient features rank second only to $q$ in influence, capturing inter-frame motion complexity. The proposed RF model achieves a MAPE = \SI{8.21}{\percent} and $R^2=0.88$, demonstrating strong predictive power.

\subsubsection{B-frame}
B-frame prediction is more challenging due to bidirectional referencing and variable motion-compensation accuracy. As depicted in Fig.~\ref{fig:shap_b_double}, temporal gradients derived from both past and future references significantly contribute to bitrate estimation, complementing QP and luma texture cues. The B-frame model achieves MAPE = \SI{12.94}{\percent} and $R^2=0.77$. Although the correlation is slightly weaker than for I- and P-frames, the overall error magnitude remains small because B-frames inherently consume fewer bits.

\subsubsection{Ablation on chroma features}
As seen in Table~\ref{tab:model_res_cons}, excluding chroma features slightly degrades performance across all frame types, with an average increase of approximately \SI{1}{}–\SI{2}{\percent} in MAPE and a drop of \SI{0.02}{}–\SI{0.04}{} in $R^2$. The effect is most noticeable in B-frames, where chroma texture provides additional information about motion-compensated prediction residuals.

\subsection{Rate control results}
\label{sec:ratecontrol_implications}
Table~\ref{tab:res_tests} compares the proposed rate control with conventional two-pass rate control (2pRC)~\cite{rc_ref, rc_ref2} on JVET UHD test sequences~\cite{seq1_ref}. Under Random Access, the proposed method achieves comparable coding efficiency, with an average BD-rate of \SI{-0.14}{\percent} versus \SI{0.26}{\percent} for 2pRC. While some sequences (\eg \textit{FoodMarket4}, \textit{ParkRunning3}) show notable gains, others exhibit marginal BD-rate increases. \vig{This behavior reflects the content-adaptive nature of the predictor, which prioritizes rate stability and bitrate conformity, occasionally trading off fractional coding efficiency on highly dynamic or irregular content. For sequences where a slight coding loss is observed, the deviation remains within the typical variability of rate-control mechanisms and is offset by substantial gains in encoding speed and rate stability, which are the primary design goals of the proposed framework.}  Overall, the method delivers competitive rate--distortion performance without re-encoding and maintains negligible deviation from the target bitrate (\SI{0.05}{\percent}), confirming robust rate stability.

Furthermore, the proposed framework significantly reduces encoding time. While the first pass of 2pRC operates at about 0.40\,fps, the model-driven pass exceeds 10\,fps, yielding a \SI{25}{\times} speedup. \vig{VCA feature extraction and Random Forest inference jointly account for less than \SI{5}{\percent} of the total encoding time.} Overall encoding time is reduced by \SI{33.33}{\percent} versus 2pRC and by \SI{-6}{\percent} compared to one-pass fixed-QP encoding. These speed and stability gains make the approach practical for real-time deployment.

\begin{table}[t]
\caption{Results compared to the fixed-QP encoding using 2pRC and the proposed method under RA configuration.}
\centering
\resizebox{0.80\linewidth}{!}{
\footnotesize
\begin{tabular}{c||c|c}
\specialrule{.12em}{.05em}{.05em}
\specialrule{.12em}{.05em}{.05em}
\multirow{2}{*}{\textbf{Sequence}} & \textbf{2pRC} & \textbf{Proposed RC} \\ 
 & $BD_{YUV}$[\%] & $BD_{YUV}$[\%] \\ 
\specialrule{.12em}{.05em}{.05em}
\specialrule{.12em}{.05em}{.05em}
Tango1	        &  0.25  &  1.12  \\
FoodMarket4	    &  0.41  &  -1.83  \\
Campfire	    &  0.18  &  0.26   \\
CatRobot        &  0.29  &  -0.64  \\
DaylightRoad2   &  0.36  &  1.47   \\
ParkRunning3    &  0.10  &  -1.22  \\
\specialrule{.12em}{.05em}{.05em}
\specialrule{.12em}{.05em}{.05em}
Average         &  \textbf{0.26}  &  \textbf{-0.14}  \\ 
\specialrule{.12em}{.05em}{.05em}
\specialrule{.12em}{.05em}{.05em}
$\Delta T$ [\%] &  \textbf{41}  &  \textbf{-6}   \\
Deviation from target rate [\%] &  \textbf{0.00}  &  \textbf{0.05}  \\ 
\specialrule{.12em}{.05em}{.05em}
\specialrule{.12em}{.05em}{.05em}
\end{tabular}}
\vspace{-0.8em}
\label{tab:res_tests}
\end{table}

\section{Conclusion}
\label{sec:conclusion}
This paper presented a content-driven framework for frame-level bit prediction in VVC, leveraging features derived from VCA. By modeling spatial and temporal characteristics aligned with encoder decision processes, the proposed Random Forest–based predictors accurately estimate bit consumption for I, P, and B frames without trial encoding. Results on UHD sequences demonstrate strong predictive accuracy ($R^2=$\SI{0.93}{}, \SI{0.88}{}, \SI{0.77} for I/P/B), confirming that lightweight VCA descriptors capture bitrate-governing complexity. Integrated into a rate-control loop under the Random Access configuration, the method delivers coding efficiency comparable to conventional two-pass rate control (2pRC): the proposed approach attains an average $BD_{YUV}$ of \SI{-0.14}{\percent} compared to \SI{0.26}{\percent} for 2pRC, while maintaining tight rate stability (deviation \SI{0.05}{\percent}). In terms of complexity, the learned first pass is about \SI{25}{\times} faster than 2pRC, and overall encoding time is reduced by \SI{33.3}{\percent} relative to 2pRC and by \SI{6}{\percent} relative to a fixed-QP anchor. The proposed approach is therefore suitable for real-time and power-constrained encoders on mobile, embedded, and cloud platforms.

Future work will extend the framework to dynamic GOP structures, enabling online model adaptation to content shifts, and integrate it into VVenC’s rate control module for energy-efficient encoding in production streaming pipelines. The framework can also support real-time encoding and adaptive bitrate streaming, where accurate per-frame bit prediction directly improves buffer stability and user experience.
\balance
\newpage
\bibliographystyle{IEEEtran}
\bibliography{poster.bib}
\balance
\end{document}